\documentclass[a4paper,fleqn,usenatbib]{mnras}

\usepackage[T1]{fontenc}
\usepackage{ae,aecompl}

\usepackage{graphicx}	\usepackage{amsmath}	\usepackage{amssymb}

	\newcommand{\ergss}{$\hspace{2pt} \rm ergs \hspace{4pt} s^{-1}$}

\title[Transient Galactic centre radio pulsar]{A transient, flat spectrum radio pulsar near the\\ Galactic centre}

\author[J. Dexter et al.]{
J. Dexter$^{1}$\thanks{E-mail: jdexter@mpe.mpg.de}, N.
Degenaar$^{2,3}$, M. Kerr$^{4}$, A. Deller$^{5,6}$, J. Deneva$^{7}$, P.
Lazarus$^{8}$,\newauthor  M. Kramer$^{8,9}$, D. Champion$^{8}$, R. Karuppusamy$^{8}$
\\
$^{1}$Max Planck Institute for Extraterrestrial Physics,
Giessenbachstr. 1, 85748 Garching, Germany\\
$^{2}$Institute of Astronomy, University of Cambridge, Madingley Road, Cambridge CB3 OHA\\
$^{3}$Anton Pannekoek Institute for Astronomy, University of
Amsterdam, Science Park 904, 1098 XH, Amsterdam, the Netherlands\\
$^{4}$CSIRO Astronomy and Space Science, Australia Telescope National Facility, Epping NSW 1710, Australia\\
$^{5}$ASTRON, the Netherlands Institute for Radio Astronomy, Postbus
2, 7990 AA, Dwingeloo, The Netherlands\\
$^{6}$Centre for Astrophysics and Supercomputing, Swinburne University
of Technology, PO Box 218, Hawthorn, VIC 3122, Australia\\
$^{7}$George Mason University, resident at the Naval Research Laboratory, Washington, DC 20375, USA\\
$^{8}$Max Planck Institute for Radio Astronomy, Auf dem H\"{u}gel 69,
53121, Bonn Germany\\
$^{9}$Jodrell Bank Centre for Astrophysics, University of Manchester, Manchester, M13 9PL, UK\\
}

\date{Accepted XXX. Received YYY; in original form ZZZ}

\pubyear{2015}

\begin{document}
\label{firstpage}
\pagerange{\pageref{firstpage}--\pageref{lastpage}}
\maketitle

\begin{abstract}
Recent studies have shown possible connections between highly
magnetized neutron stars (``magnetars''), whose X-ray emission is too
bright to be powered by rotational energy, and ordinary radio
pulsars. In addition to the magnetar SGR J1745$-$2900, one of the radio
pulsars in the Galactic centre (GC) region, PSR J1746$-$2850, had
timing properties implying a large magnetic field strength and young age,
as well as a flat spectrum. All characteristics are similar to those of rare, transient,
radio-loud magnetars. Using several deep non-detections from the
literature and two new detections, we show that this pulsar is also transient in the
radio. Both the flat spectrum and large amplitude variability are inconsistent with the
light curves and spectral indices of 3 radio pulsars with high
magnetic field strengths. We further use frequent, deep archival imaging observations of
the GC in the past 15 years to rule out a possible X-ray outburst with
a luminosity exceeding the rotational spin down rate. This source, either a transient magnetar without any
detected X-ray counterpart or a young, strongly magnetized radio pulsar producing
magnetar-like radio emission, further blurs the line
between the two categories. We discuss the 
implications of this object for the radio emission mechanism in
magnetars and for star and compact object formation in the GC.
\end{abstract}

\begin{keywords}
pulsars: individual (PSR J1746$-$2850) --- stars: magnetars --- Galaxy: centre
\end{keywords}

\section{Introduction}

The pulsed radio emission from neutron stars (pulsars) is
thought to be powered by their rotation. In some rare, young neutron
stars with high magnetic field strengths \citep[anamalous X-ray pulsars,
AXPs, and soft-gamma repeaters, SGRs, $B \sim
B_{\rm crit} \gtrsim 4.4\times10^{13}$ G, e.g.,][]{turollaetal2015}, the observed
X-ray luminosity can exceed the magnetic dipole spin down
luminosity. According to the
``magnetar'' model, this X-ray emission is instead powered by
dissipation of the magnetic field
\citep{thompsonetal2002}. Observationally
  \citep[e.g.,][]{reaetal2014}, magnetars have relatively
large spin periods ($P \simeq 0.3-12$ s) and period derivatives
($\dot{P} \sim 10^{-15} - 10^{-10}$ s/s), typical X-ray luminosities
of $L_X \sim 10^{31-35}$ erg / s, and experience episodes of enhanced
X-ray activity that can be either a long-lived "outburst" (lasting
months to years; transient magnetars) or short-lived
bursts and flares (lasting seconds to minutes).

The discoveries of radio pulsars with high
inferred dipolar magnetic fields
\citep[high-B pulsars,][]{camiloetal2000,mclaughlinetal2003}, of magnetars with low
dipolar field strength $B < B_{\rm crit}$ \citep[e.g.,][]{reaetal2010}, and of
transient pulsed radio emission from magnetars \citep{camiloetal2006}
show that these sources are not separated solely by the inferred dipolar magnetic
field strength. The recent discovery of X-ray outbursts from a high-B
radio pulsar suggests that objects can be hybrids between classes
\citep{gogusetal2016,archibaldetal2016}. The pulsed radio emission
from $4$ \citep[PSR J1622$-$4950, 1E 1547.0$-$5408, XTE J1810$-$197,
SGR J1745$-$2900,][]{camiloetal2006,camiloetal2008,levinetal2010,eatoughetal2013}
out of $28$\footnote{\url{http://www.physics.mcgill.ca/~pulsar/magnetar/main.html}}
magnetars (all transient) is associated with X-ray outbursts
  \citep{burgayetal2006,crawfordetal2007,lazarusetal2012}. It provides important clues to the
outburst mechanism and magnetospheric structure, but remains poorly
understood. This radio emission is distinct from that of ordinary
radio pulsars, with a flat spectrum \citep[e.g.,][]{camiloetal2008} and large amplitude flux
variability \citep[e.g.,][]{lazaridisetal2008}. All known radio-loud magnetars have low quiescent X-ray
luminosities below the spin down rate \citep{reaetal2012}.

The radio pulsar PSR J1746$-$2850 \citep{denevaetal2009detect}, one of six known
pulsars within $15$ arcmin ($\simeq 25$ pc in projected distance) of Sgr A* in the
Galactic centre (GC), was found to have timing properties ($P \simeq
1.1$ s, $\dot{P} \simeq 1.3\times10^{-12}$ s/s) implying a young age ($T \simeq 13$ kyr) and a high
magnetic field strength ($\simeq 4.2\times10^{13}$ G), as well as a flat radio spectrum
($\alpha \simeq -0.3$). All properties are characteristic of transient magnetars 
\citep[e.g.,][]{camiloetal2006,torneetal2015}. Using archival radio data and reported limits from the literature, we
compile a light curve of PSR J1746$-$2850 to show that it is a transient
radio pulsar (\S \ref{sec:j1746-2850-radio}). We then analyse densely
sampled \emph{XMM-Newton} archival data of the GC to place limits on the X-ray luminosity going
back to 2000 (\S \ref{sec:j1746-2850-x}) and a deep archival
\emph{Chandra} image to place a deep limit on its quiescent X-ray
luminosity. We further show that the spectral indices and light curves of 3
high-B pulsars are similar to those of the ordinary pulsar population
and inconsistent with those of PSR J1746$-$2850 (\S \ref{sec:j1746-2850-as}). We
argue that either PSR J1746$-$2850 is a transient magnetar, in which case it would be the second to be detected from its
radio emission and the only one for which no X-ray emission has been
detected to date, or a new hybrid with magnetar-like radio emission but no
X-ray outbursts. This object further blurs the
line between magnetars and ordinary radio 
pulsars, and we discuss its implications for the magnetar radio emission
mechanism and for future high-frequency pulsar searches of the GC (\S
\ref{sec:discussion}).

\begin{figure*}
\begin{center}
\includegraphics[scale=1.0]{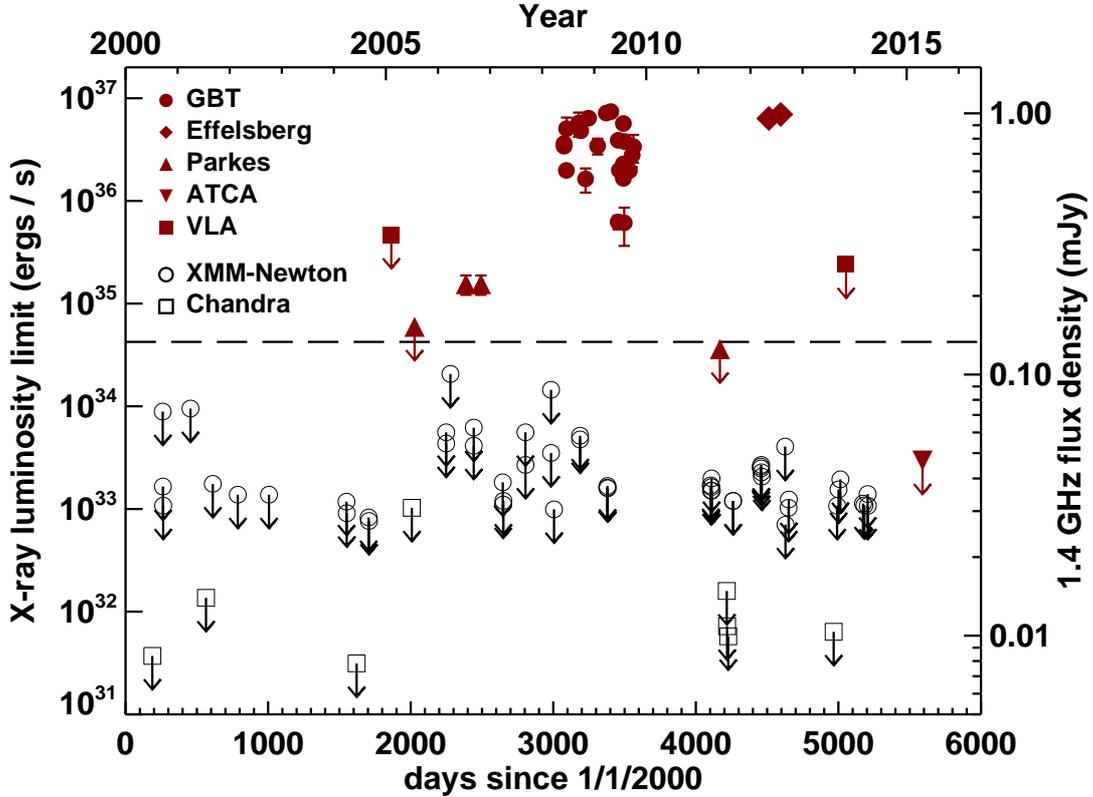}
\caption{\label{lightcurve}Radio light curve of PSR J1746$-$2850 from pulsar timing and
  searches (solid points, right axis) and imaging (squares), as
  well as 
  X-ray non-detections (open points, left axis). The
  pulsar was originally detected and timed in GBT radio observations in
  2008 and 2009 \citep{denevaetal2009detect}, and was later also 
found in Parkes survey data from 2006 \citep{batesetal2011}. It
was not detected in a previous Parkes survey in
2005 \citep{johnstonetal2006} or in 2011 \citep{ngetal2015}, but was
then detected at a high flux density in 2012 with Effelsberg. Since then there have
been upper limits from VLA images \citep{ludovicietal2016} and gated
ATCA images \citep{schnitzeleretal2016}. PSR J1746$-$2850 has not been detected in the X-ray, despite
dense, deep imaging of the GC region since 2000, leading to many upper
limits (open points). The limits have been converted to unabsorbed
X-ray luminosity assuming a GC distance of $8.3$ kpc
\citep[e.g.,][]{chatzopoulosetal2015,gillessenetal2017} and using the column density ($N_H
\simeq 1.9 \times 10^{23} \hspace{2pt} \rm cm^{-2}$)
and spectrum ($k T_{\rm bb} \simeq 1$ keV) towards the nearby GC magnetar SGR J1745$-$2900
\citep{cotizelatietal2015}. All limits rule out a
a magnetar outburst in the X-rays ($L_X > \dot{E} \simeq 4.2\times10^{34}$ \ergss, long dashed line).}
\end{center}
\end{figure*}

\begin{figure}
\begin{center}
\includegraphics[scale=0.8]{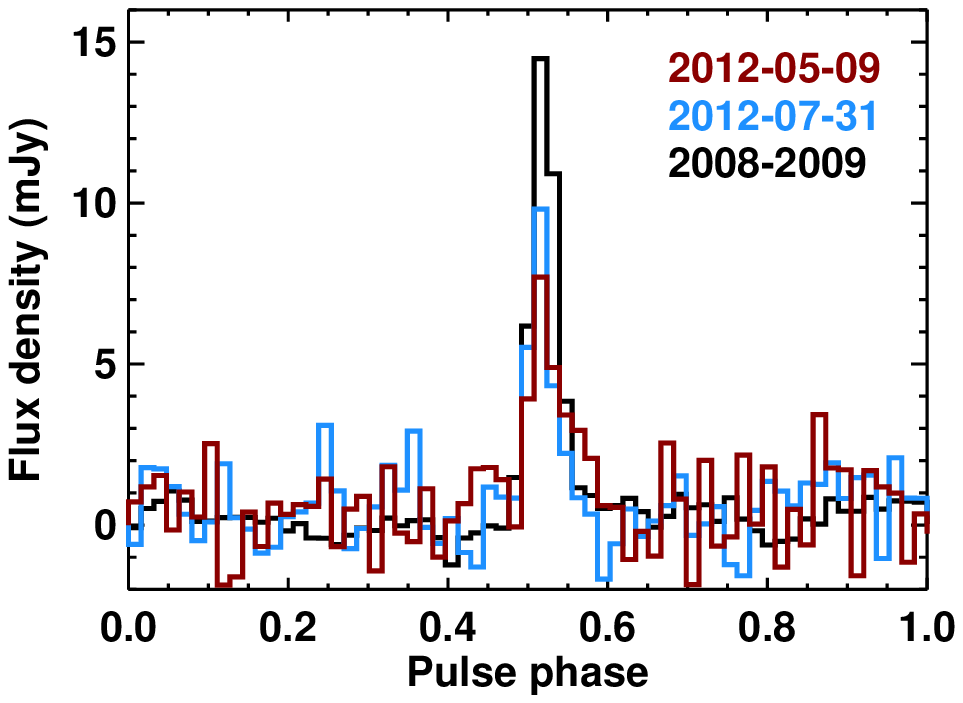}
\caption{\label{lazarus}Pulse profiles of PSR J1746$-$2850 measured at
  Effelsberg in May and July 2012 (red and blue lines) compared with an
  averaged 2 GHz profile \citep[black line,][]{denevaetal2009detect}. The pulsar was detected with a flux
  density comparable to that of its peak during 2008-2009, despite
  multiple, deep non-detections before and since.}
\end{center}
\end{figure}

\section{PSR J1746$-$2850 radio emission and recent non-detections}
\label{sec:j1746-2850-radio}

We compile a radio light curve (solid points and upper limits, right
axis of figure \ref{lightcurve}) of PSR J1746$-$2850 using a mix of
archival and new data and published results. We estimate flux
densities from the folded pulse profile data from the timing
observations reported in \citet{denevaetal2009detect}, using the observed signal-to-noise ratio
and the estimated noise at the GBT. We have removed epochs that were
clearly affected by radio frequency interference, leaving observations
at $2.0$, $4.8$, and $9.0$ GHz. From those data, we confirm the
\citet{denevaetal2009detect} spectral index measurement of $\alpha =
-0.3$ (see \S \ref{sec:j1746-2850-as} for details), and use that to
scale all flux densities to $1.4$ GHz in figure \ref{lightcurve}. Some
RFI contamination is still present and so the intrinsic source
variability is likely less than the scatter in the data.

Figure \ref{lazarus} shows the average pulse profiles from
  detections of PSR J1746$-$2850 
in two previously unpublished observations at $\nu \simeq 2.7$ GHz from Effelsberg in May and
July 2012 using the new PSRIX backend \citep[see ][for more
details]{lazarusetal2016}, compared with the best profile from GBT observations. The data were flux calibrated using
observations of on- and off-source scans of radio sources with known,
stable flux densities. From the calibrated pulse
profiles, we estimate average flux densities of $0.79 \pm 0.18$ and
$0.82 \pm 0.16$ mJy, comparable to the brightest flux densities seen
with the GBT (two latest detections in figure \ref{lightcurve}) and
with a very similar pulse profile.

Next we add flux densities and upper limits from the literature, also shown
scaled to $1.4$ GHz. The earliest detection of PSR J1746$-$2850 was in
October 2006 from a Parkes
multibeam survey \citep{batesetal2011}. It was not found in a previous
Galactic Center pulsar survey \citep{johnstonetal2006} from July
2005. In June 2011, the pulsar was not detected using Parkes \citep{ngetal2015}. PSR
J1746-2850 should have been seen by both of these surveys if it was
active at the time: even the lowest flux density seen by
\citet{batesetal2011} would have provided a signal--to--noise ratio of
$\sim$15. \citet{schnitzeleretal2016} observed this source with gated ATCA
observations in 2015. The $6\sigma$ peak flux
density limit was $< 0.5$ mJy at
$5$ GHz. \citep{schnitzeleretal2016}, a limit 
$\simeq 15-30\times$ smaller than the average from $2008-2009$ and
$2012$ (figure \ref{lightcurve}). We scale this peak flux density limit to an average flux
density $< 0.04$ mJy by assuming the 2012 pulse profile (deep, recent radio upper limit in
figure \ref{lightcurve}). There have been further non-detections
since 2013 with the GBT (Siemion et al., priv. comm.), ATCA,
Effelsberg, and Parkes, both from folded and search mode data. These are not shown in figure \ref{lightcurve} because
observation epochs and/or flux density limits are not available.

Since the original timing solution was successfully used to
detect the pulsar in 2012, it seems unlikely that the non-detections
could be due to timing irregularities. Another possible explanation for non-detections in timing observations
is time-variable scattering. All pulsars observed to date in the GC
region exhibit large amounts of temporal broadening 
\citep{johnstonetal2006,denevaetal2009detect,spitleretal2014}, $\tau
\simeq 1-3$s scaled to $1$ GHz assuming $\tau \propto \nu^{-4}$. When
$\tau \simeq P$, the pulse is smeared and becomes 
difficult to detect with time domain or gated observations. However, the upper
limits discussed for PSR J1746$-$2850 are from 
observations at $3.1-5$ GHz (or even $> 10$ GHz, A. Siemion,
priv. comm.), where this degree of scattering is negligible. Further,
the source has been detected over the full range $1.4-8.4$ GHz at
multiple epochs. There is therefore no evidence for time variability in the
scattering medium that could produce a degree of temporal broadening $\sim
2$ orders of magnitude larger.

To further check these possible explanations, we looked for
PSR J1746$-$2850 in deep VLA images. Two archival, high resolution
observations covering the location of PSR J1746$-$2850 were available,
taken in 2005 and 2013 (see figure \ref{lightcurve}). The Feb 2005
data were taken at $8.4$ GHz with 100 MHz of total bandwidth and $2.2$
hr  integration time in AB configuration. No point source was
identified within $2 \sigma$ of the timing position of PSR
J1746$-$2850 ($10\times6$ arcsec). The highest peak within this region was $\simeq 0.14$
mJy/beam, and we place a $3 \sigma$ upper limit of $0.21$ mJy at 8.4
GHz. The 2013 observations covered multiple VLA configurations from
DnC to BnA, with an instantaneous bandwidth of 2 GHz centred on 5 GHz
\citep{ludovicietal2016}. Again no point source was found near the
timing position of PSR J1746$-$2850. The highest peak of 
the image was $0.12$ mJy/beam within this region, leading to a $3 \sigma$ upper limit of $0.18$
mJy. The point source sensitivity in both images is limited by significant diffuse
emission in the region. Nonetheless, the limits are factors of $\simeq
1.5$ and $2$ lower than the average flux densities measured with the
GBT at $5$ and $8.4$ GHz. There could be additional systematic errors
on the pulsar timing position, and there are some what higher peaks
nearby. However, these appear to be from diffuse emission. While the imaging limits are not as deep as those
from timing, they cannot be affected by uncertainties in the timing
solution. Since it is the most likely
explanation, we conclude that the non-detections are due to intrinsic
source variability and that PSR J1746$-$2850 is a transient pulsar.

\begin{figure*}
\begin{center}
\includegraphics[scale=0.5]{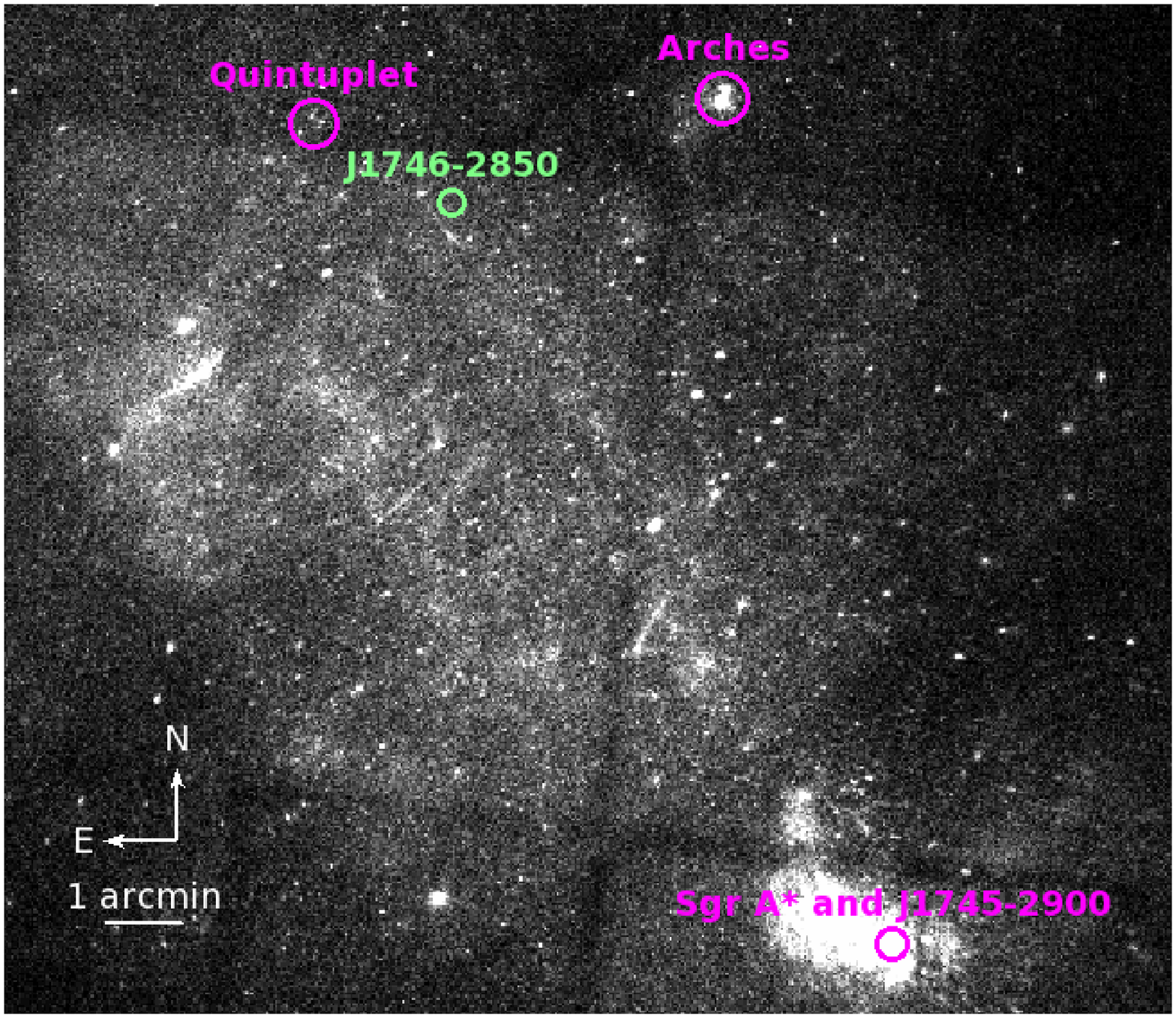}
\caption{\label{chandraimage}Composite \emph{Chandra} ACIS-I $392$ ksec image of the
  GC, with the radio location of PSR J1746$-$2850 and
  its $5 \sigma$ uncertainty marked by the green circle (towards top left). No source is detected
  near the expected position, leading to stringent limits on its absorbed
  X-ray luminosity $\lesssim 10^{30}$ \ergss (but weaker
  limits on the intrinsic unabsorbed 
 bolometric luminosity for a thermal spectrum, see Table
 \ref{quiescentxray}). The pulsar could have been produced in
 the nearby, young Arches or Quintuplet clusters
 \citep{denevaetal2009detect}. The positions of Sgr A* and SGR
 J1745-2900 are shown for comparison.}
\end{center}
\end{figure*}

\begin{table}
\caption{\emph{Chandra} upper limits on the quiescent 
  temperature and luminosity of PSR J1746$-$2850 for assumed values of $R$, $N_H$\label{quiescentxray}}
\begin{minipage}{6cm}
\begin{small}
\begin{center}
\begin{tabular}{llll}
        \hline
$R$ (km) & $N_H$ ($10^{22} \rm \hspace{2pt} cm^{-2}$) & $k T_{bb}$
(keV) & $L_{\rm bol}$
($10^{33}$ \ergss)\\
        \hline
$10$ & 1.0 & 0.08 & 0.5\\
$10$ & 19 & 0.18 & 14\\
$1$ & 1.0 &  0.12& 0.03\\
$1$ & 19 & 0.27 & 0.7\\
	\hline
\end{tabular}
\end{center}
\end{small}
\end{minipage}
\end{table}

\section{Archival X-ray data}
\label{sec:j1746-2850-x}

We looked for an X-ray counterpart to PSR J1746$-$2850 in archival 
images of the GC. The position is in the field of view of
\emph{XMM-Newton}/EPIC imaging observations of Sgr A*, providing a
large number of observations from 2000 to the present, totaling
$\simeq 2.4$ Msec of exposure time. Almost all images were obtained
with the PN camera, while a handful were from MOS1. No source was detected within $< 20$
arcsec ($\gtrsim 3\sigma$) of the position of PSR J1746$-$2850 in any of the images. We used the default pipeline output
sensitivity maps to obtain upper limits on the observed source
flux. These values were then converted to unabsorbed (intrinsic)
luminosity limits (open circles in figure \ref{lightcurve}) using the column density ($N_H
\simeq 1.9\times10^{23} \rm cm^{-2}$) and spectrum (blackbody with $k
T \simeq 1$ keV) of the GC magnetar SGR J1745$-$2900
\citep{reaetal2013,cotizelatietal2015}, and a
distance of $8.3$ kpc to the GC \citep[e.g.,][]{chatzopoulosetal2015,gillessenetal2017}. Accounting
for the spectral shape and absorption leads to a correction by factors of $2.5-5$ depending on the camera
and filter used for each observation.

The radio position of PSR J1746$-$2850 is not within the field of view for
\emph{Chandra} when centered on Sgr A*, but its position was still
observed in offset pointings at several epochs with the ACIS-I
instrument, totaling $392$ ksec of exposure time. No counts were
observed from near this source position. The closest point sources
detected by \citet{munoetal2009} are $\gtrsim 20$ arcsec away,
$\gtrsim 3 \sigma$ from the pulsar timing position and far enough that
the false positive rate for finding an X-ray point source becomes high. We estimate upper limits by
conservatively assuming 3 photons at the source position
\citep[e.g.,][]{gehrels1986}. We converted the count rate limits to
unabsorbed luminosity limits (open squares in figure \ref{lightcurve})
for the same parameters as above. For ACIS-I the correction for the
expected spectrum and column density is a
factor $\simeq 2.5$. 

The \emph{Chandra} data provide much deeper limits than 
\emph{XMM} due to the very low instrumental background. We created a combined image (figure \ref{chandraimage})
and used this long exposure to obtain a limit of
$1.2\times10^{-5} \hspace{2pt} \rm counts / \rm s$. Neglecting
interstellar absorption, this corresponds to a quiescent luminosity limit of $L_X <
10^{30}$ \ergss. We convert this into an intrinsic
neutron star temperature by assuming a blackbody spectrum along with
fiducial values for the GC column
density and neutron star emission radius. These are $N_H = 19$, $1 \times 10^{22} \rm cm^{-2}$ as observed towards
SGR J1745$-$2900 and towards lower absorption regions of the GC
\citep{munoetal2009}, and $R = 1$, $10$ km corresponding to the polar
cap or neutron star surface. The results are given in Table
\ref{quiescentxray}. The limits on intrinsic, quiescent luminosity are $\lesssim
10^{33}$ \ergss, lower than or comparable to the measured values for
many transient magnetars and high-B pulsars \citep[e.g.,][]{reaetal2012}.

\begin{figure*}
\begin{center}
\begin{tabular}{cc}
\includegraphics[scale=0.67]{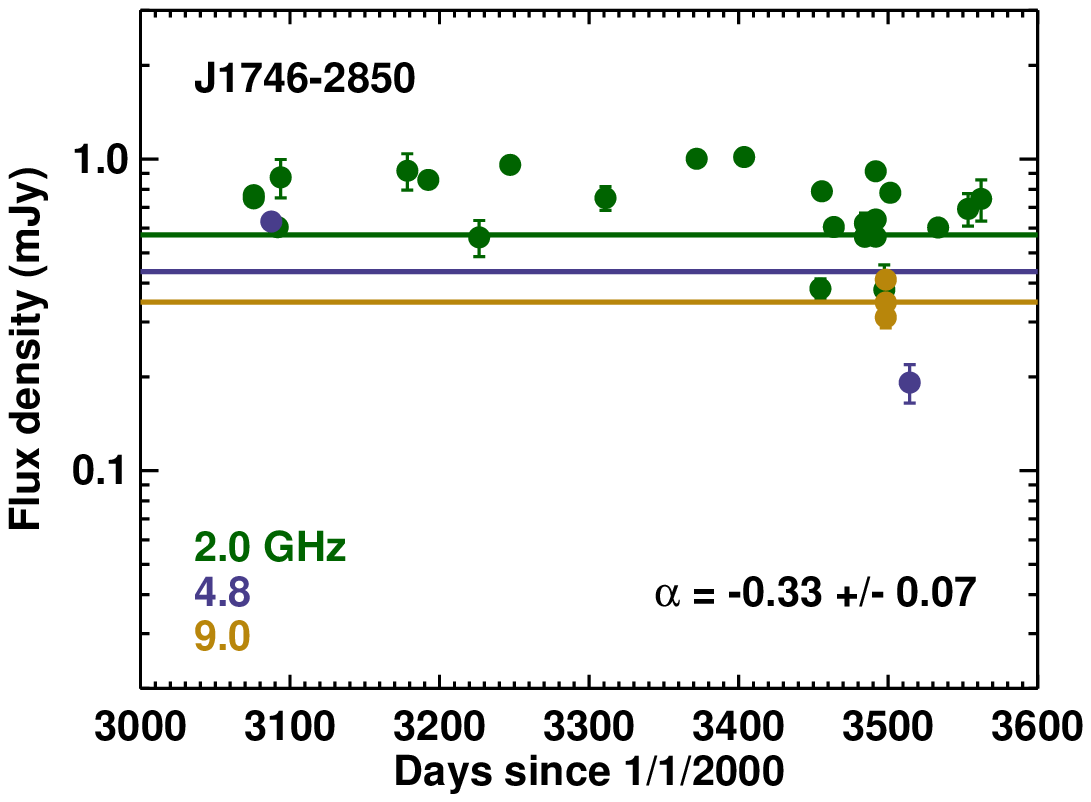}&
\includegraphics[scale=0.67]{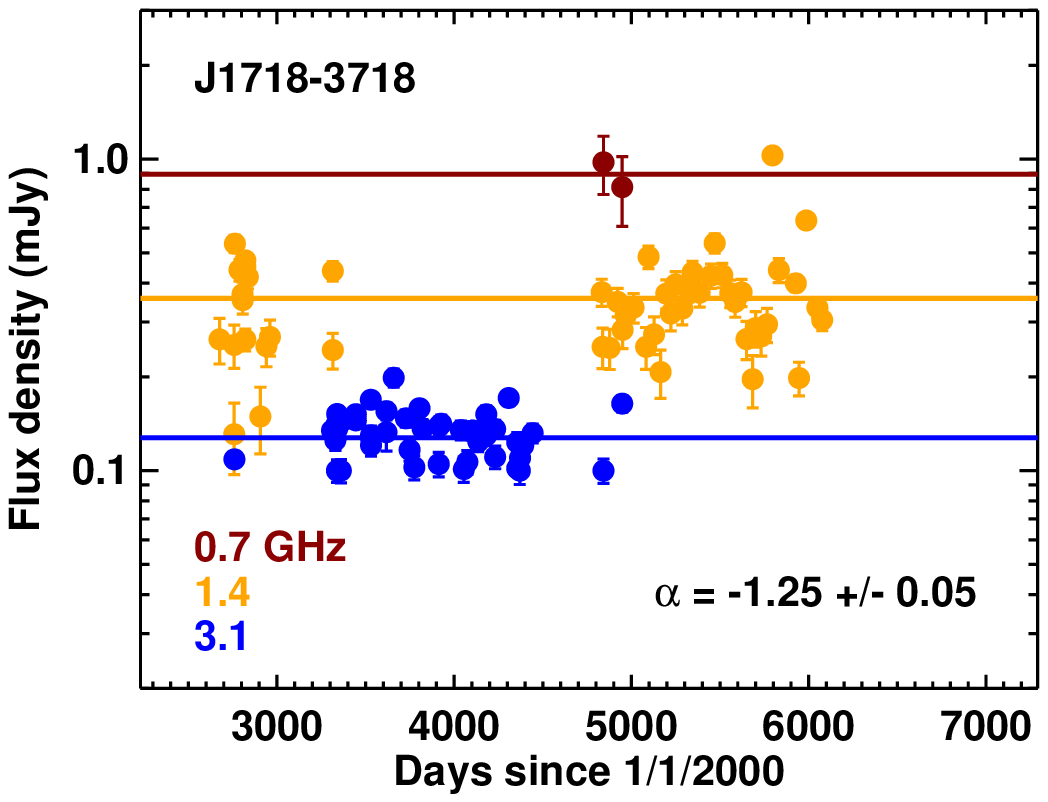}\\
\includegraphics[scale=0.67]{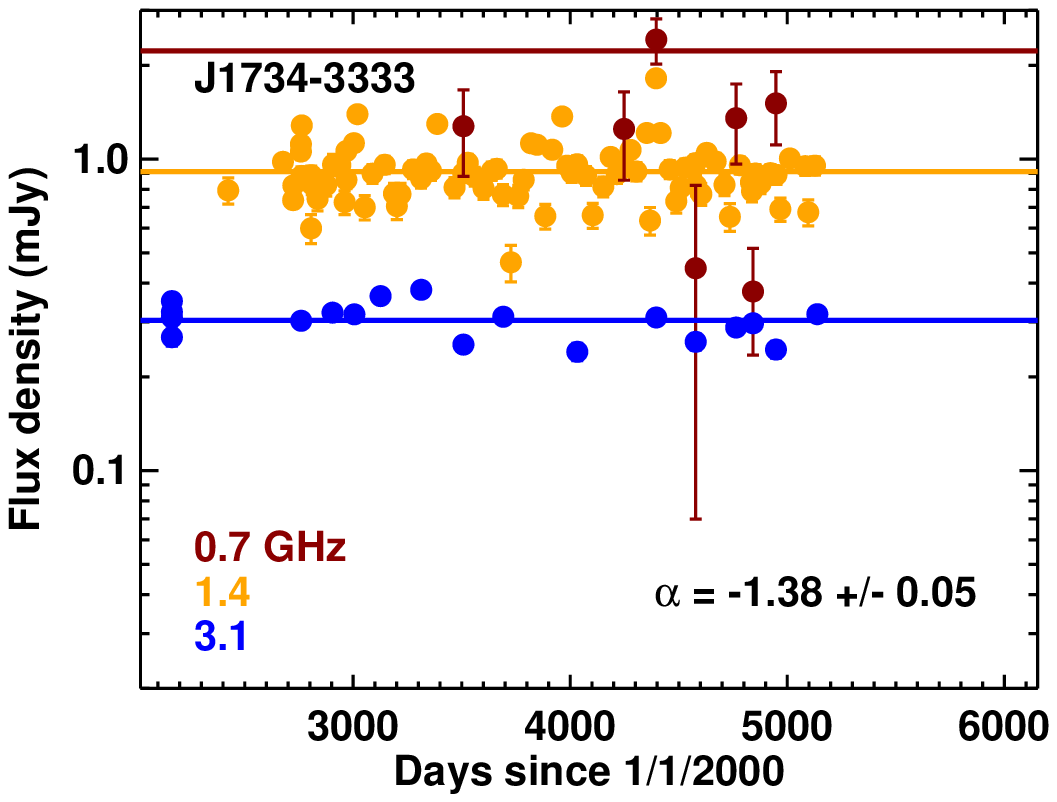}&
\includegraphics[scale=0.67]{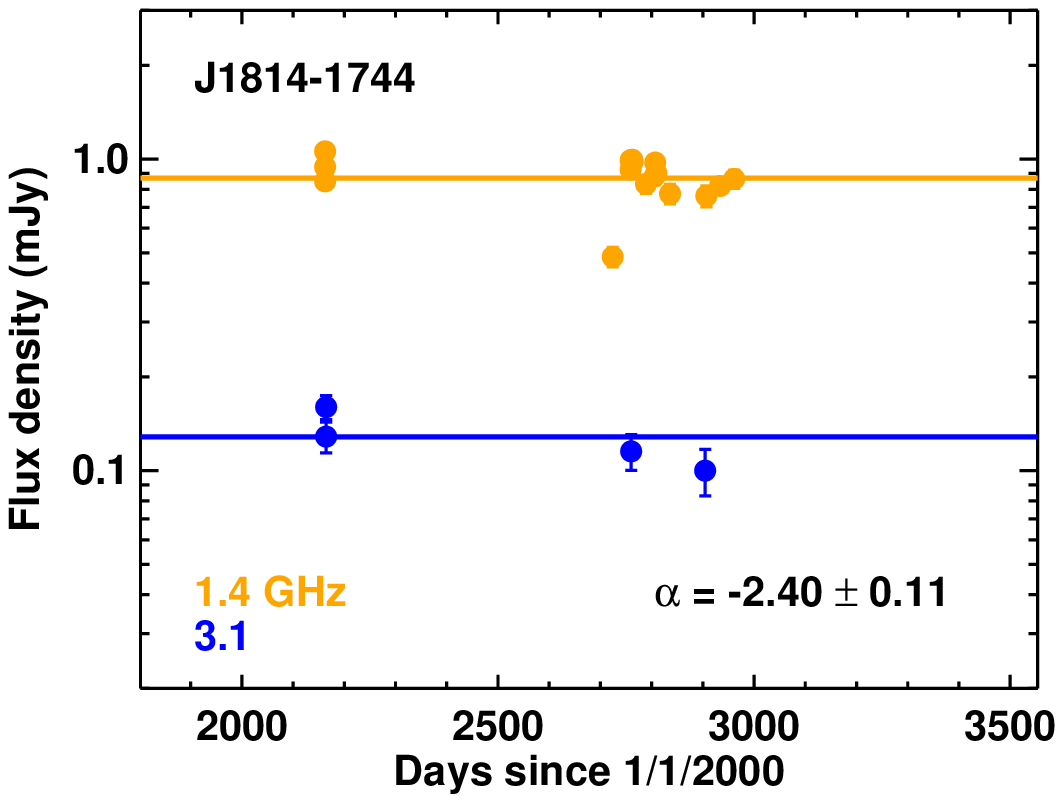}
\end{tabular}
\caption{\label{specindex}Light curves from timing observations of
  PSR J1746$-$2850 with the GBT in 2008-2009 \citep[top left,][]{denevaetal2009detect} and three high-B
  pulsars (other panels) with Parkes. The radio spectrum (measured
  from the averages at each frequency, horizontal lines) of PSR J1746$-$2850 was flat ($\alpha
  \simeq -0.3$), as seen in radio-loud magnetars. The other high-B
  pulsars have steep radio spectra ($\alpha \sim -1.2$ to $-2.5$),
  within the normal range for ordinary pulsars,
  and have been detected consistently over several years unlike PSR
  J1746$-$2850 (see figure \ref{lightcurve}).}
\end{center}
\end{figure*}

\section{Extreme pulsar variability}

As described above, the radio emission from PSR J1746$-$2850 appears to
be transient. In addition to radio loud magnetars, the known classes
of radio pulsars with high amplitude variability on long timescales include transitional
millisecond pulsars, which switch between pulsed
radio and X-ray emission \citep[e.g.,][]{archibaldetal2009,papittoetal2013} and intermittent pulsars
\citep{krameretal2006}. Since PSR J1746$-$2850 has a long period $\simeq
1$s and no detected X-ray emission, we exclude the former
possibility.

Intermittent pulsars are objects with 2-3 distinct states of radio
emission with large differences in flux density. The duty cycles
for the bright states can be short \citep[$\simeq
10-20\%$, e.g.,][]{youngetal2013}, and a wide range of intermittency periods are
seen. Hence, the radio light curve of 
PSR J1746$-$2850 would be roughly compatible with ``on'' and ``off''
periods of duration $\simeq 1$ year, similar to those of the
intermittent pulsars PSR J1832+0029 \citep{lorimeretal2012} and
J1841$-$0500 \citep{camiloetal2012}.

We disfavor this interpretation, because other than their transience,
the intermittent pulsars discovered to date have characteristic ages,
magnetic field strengths, and radio spectra comparable to normal radio
pulsars \citep[e.g.,][]{lyneetal2016}. As a young, high magnetic field
object with a flat radio spectrum, PSR J1746$-$2850 differs in all three
of these diagnostics.

\section{PSR J1746$-$2850: a transient, flat spectrum, high-B radio pulsar}
\label{sec:j1746-2850-as}

The large period and period derivative of PSR J1746$-$2850 \citep{denevaetal2009detect} imply a high surface
magnetic field strength, young age, and high spin down luminosity,
which classify it either as a high-B pulsar \citep{camiloetal2000}, or
a transient magnetar (SGR source). We have further shown that this
radio emission is transient and has a flat spectrum, similar to that
seen in the $4$ radio-loud magnetars.

The X-ray constraints over many epochs described in \S2
rule out steady magnetar activity and constrain
the duration of any magnetar outbursts with $L_X > \dot{E} \simeq
4 \times 10^{34}$ \ergss $\hspace{2pt}$ to several months, where
$\dot{E} \propto \dot{P} / P^3$ is the magnetic dipole spin down luminosity. X-ray
outbursts with roughly exponential decay profiles have been found to precede the radio emission from
magnetars \citep[e.g.,][]{reaesposito2011}. In order to trigger the
radio detections in 2006, 2008-2009, 
and/or 2012 and not have been detected, an X-ray outburst would have had to have a decay time $\tau <
100$ d for any possible start time. This is shorter than usual
outbursts in radio-loud magnetars: $\tau \sim 100-600$ days
\citep{gavriiletal2004,gotthelfhalpern2007,andersonetal2012,cotizelatietal2015},
although the magnetar 1E 1547.0$-$5408 has shown X-ray outbursts that fade on
few week timescales \citep{israeletal2010}. The defining feature of
magnetars is X-ray emission exceeding the spin down luminosity, and so
the absence of any X-ray detection and strict limits on even one possible outburst preceding the radio
detections in 2006, 2008-2009, or 2012 prevent us from conclusively classifying PSR J1746$-$2850
as a radio-loud, transient magnetar.

We then ask whether other known high-B pulsars show transient, flat
spectrum radio emission. We measured light curves and spectra for
three pulsars with surface dipolar $B \simeq 3.2 \times 10^{19}
\sqrt{P \dot{P}} \hspace{2pt} \rm G >
2 \times 10^{13} \hspace{2pt} \rm G$ observed at 1.4 and 3.1 GHz from Parkes monitoring
observations (figure \ref{specindex}). The 
spectral indices range from -2.5 to -1.2, similar to those of ordinary
radio pulsars \citep{lorimeretal1995,batesetal2013} and significantly
steeper than the $\alpha \simeq -0.3$ of PSR J1746$-$2850 (top left
panel of figure \ref{specindex}). The sources in this sample
vary in radio flux density by factors of a few over $\simeq 8$ years, similar
to PSR J1746$-$2850 during 2008-2009. However, none were found to drop by more
than an order of magnitude in flux for months or years, as is required
to explain the many radio non-detections of PSR J1746$-$2850 (figure
\ref{lightcurve}).

Each of the four known radio-loud magnetars, on the other hand,
produces transient, flat spectrum radio emission with $\alpha \sim 0$
\citep[e.g.,][]{camiloetal20071810,camiloetal2008,levinetal2012,torneetal2015}. this mode of radio emission is ``magnetar-like'' and not commonly seen
in high-B pulsars. Its magnetar-like radio emission and timing properties 
(large $P$ and $\dot{P}$) make PSR J1746$-$2850 a candidate
radio-loud magnetar, even without an X-ray detection. SGR J1622$-$4950 was initially detected in X-ray
quiescence \citep{levinetal2010} before being shown to have had an apparent X-ray outburst 2-3
 years earlier \citep{andersonetal2012}. The radio emission of SGR
J1745$-$2900 has shown no signs of fading after a few years, and has already
outlasted the X-ray outburst. It is therefore possible that we missed
a long ago or short duration X-ray outburst from PSR J1746$-$2850,
and/or that an outburst will be seen in the future.

Either way, PSR J1746$-$2850 is a transient, flat spectrum, high-B
radio pulsar. The magnetar-like radio emission and lack of X-ray outbursts place it as a hybrid
source between magnetars and high-B pulsars. Along with recent findings of
magnetar X-ray outbursts from a pulsar with a relatively low surface
dipole field strength \citep{reaetal2010} and an otherwise ordinary high-B
radio pulsar \citep{archibaldetal2016,gogusetal2016}, our findings suggest that there is
no distinct boundary between transient magnetars and the
high-B end of the ordinary pulsar population.

\section{Discussion}
\label{sec:discussion}

PSR J1746$-$2850 is the fifth object to show magnetar-like flat spectrum
radio outbursts, the second to be discovered in the
radio, and the only one with no detected X-ray emission (quiescent or
outburst) thus far. It is either an ``X-ray quiet'' transient magnetar
or a high-B pulsar with magnetar-like radio emission. This finding has implications for the
connections between high-B pulsars and magnetars, the magnetospheric physics of
neutron stars, and for star formation, pulsar populations, and the
detection of addititional flat spectrum radio pulsars near the GC. 

\subsection{Magnetar emission mechanism}

The leading paradigm for magnetar outbursts posits that they are produced by dissipation of
magnetospheric energy caused by motions of the neutron star crust \citep[e.g.,][]{thompsonetal2002}. The
deep, densely sampled non-detections of an X-ray outburst 
from PSR J1746$-$2850 provide further evidence that the radio emission from
transient magnetars can be produced independent of the X-ray outburst. Evidently it
may either persist many years after an X-ray outburst, or may require only a
very weak or short duration outburst (or none at all). If the radio emission is triggered due to a re-arrangement of the
structure of the magnetosphere following an outburst, either the timescale
for further evolution should be long or the 
re-arrangement should require only a small amount of dissipated energy.

Like the GC magnetar SGR J1745$-$2900, PSR J1746$-$2850 has a low
observed $L_{X, \rm qui} / \dot{E} <
10^{-2}$. Depending on the column and emission radius, the resulting
limit on the surface
temperature is as weak as $kT < 0.3$ keV (Table \ref{quiescentxray}),
but nonetheless is another example where the quiescent X-ray emission from
high-B pulsars / transient magnetars can be consistent with that of ordinary pulsars, unlike
the persistent high X-ray luminosities seen in many magnetars
(especially AXPs). 

\subsection{GC star formation}

The GC region has a large star formation and supernova rate, implying
a large birth rate of compact objects and radio pulsars. To date only
6 have been detected, including the GC magnetar SGR
J1745$-$2900. Young, high-B objects showing magnetar-like radio outbursts
make up 2/6 of the  known pulsars, suggesting either efficient
magnetar formation in the GC or selection effects biasing surveys
against the detection of ordinary, steep spectrum radio pulsar emission.

A radio proper motion study of SGR J1745$-$2900 showed that is is likely
associated with the young stars in the central parsec of the Galaxy
\citep{boweretal2015}. The origin of the other GC pulsars, including
PSR J1746$-$2850, is less clear. Given its young age, PSR J1746$-$2850 may be
associated with the Quintuplet (or possibly the 
Arches) cluster (figure \ref{chandraimage}). For a typical neutron star kick velocity, PSR J1746$-$2850
could have reached its radio position in its short lifetime
\citep{denevaetal2009detect}. Alternatively, the progenitor star
itself could have been stripped from one of the clusters and would have had
more time to travel to the observed radio position
\citep{habibietal2014}. If the source again re-brightens in the radio, its
proper motion should be measurable and could distinguish between these
scenarios.

An association with one of these clusters would be additional evidence
that they are old enough to produce neutron stars
\citep{figeretal1999}, as well as an additional association
of a high-B pulsar or magnetar with young and massive stellar populations
\citep[e.g.,][]{gaensleretal2005,munoetal2006}. 

\subsection{Predictions}

SGR J1622$-$4950 was first detected from its flat spectrum radio
emission, potentially similar to PSR J1746$-$2850. Two years later it
underwent an X-ray outburst and was confirmed to be a transient
magnetar. Future X-ray observations could find a magnetar outburst
from PSR J1746$-$2850. Further radio observations may also find a
further re-brightening or a change of spectral slope that might point
to connections with other known high-B pulsars. We therefore encourage
further monitoring of the source in the radio and X-ray.

SGR J1745$-$2900 showed a similarly flat radio spectrum as PSR J1746$-$2850, and has
now been detected up to high frequency
\citep[$\nu = 291$ GHz,][]{torneetal2016}. Future high sensitivity observations with IRAM,
the LMT, and/or phased ALMA could constrain the population of neutron
stars with magnetar-like radio emission, providing a new method for finding
these rare objects.

\section*{Acknowledgements}
We thank C. Lang and D. Ludovici for sharing their reduced, high resolution VLA data
covering the position of PSR J1746$-$2850, and G. Desvignes,
R. Eatough, P. Esposito, D. Schnitzeler, and A. Siemion for useful discussions. J. Dexter was supported by a Sofja
Kovalevskaja Award from the Alexander von Humboldt Foundation of
Germany. N. Degenaar acknowledges support via a Vidi grant from the
Netherlands Organization for Scientific research (NWO) and a 
Marie Curie fellowship from the European Commission under contract
no. FP-PEOPLE-2013-IEF-627148. M. Kramer and R. Karuppusamy acknowledge the
financial support by the European Research Council for the ERC Synergy
Grant BlackHoleCam under contract no. 610058. The Parkes radio telescope is part of
the Australia Telescope, which is funded by the Commonwealth
Government for operation as a National Facility managed by CSIRO.

\bibliographystyle{mnras}
\bibliography{pulsars}

\bsp	\label{lastpage}
\end{document}